\documentclass[journal]{IEEEtran}
\IEEEoverridecommandlockouts
\usepackage{times,amsmath,color,amssymb,graphicx,epsfig,cite,psfrag,subfigure,algorithm,balance}
\usepackage{amsfonts,pifont,enumerate,cases}
\usepackage{mathrsfs} % \mathcal ºê°üÖ§³Ö
\usepackage[table]{xcolor} % ÓÃÓÚ±í¸ñ¸÷ÐÐÑÕÉ«
\usepackage{verbatim} % Óë\begin{comment}...\end{comment}ÅäºÏÓÃ»§×¢ÊÍ
\usepackage{bm}
\usepackage{cuted,stfloats}
\usepackage{algorithm}
\usepackage{algorithmic}

\usepackage{longtable}
\usepackage{blindtext}
\usepackage{multirow}
\usepackage{float}
\usepackage{threeparttable}
\usepackage{makecell}
\usepackage[utf8]{inputenc}
\usepackage{url}
\usepackage{booktabs}
\usepackage{amssymb}
\usepackage{bbding}
\usepackage{pifont}
\usepackage{wasysym}
\usepackage{utfsym}
\usepackage{fontawesome}
\usepackage[algo2e,ruled,vlined,linesnumbered,lined,boxed,commentsnumbered]{algorithm2e}
\usepackage{amsmath,mathtools}
\usepackage[
    colorlinks=true,
    linkcolor=blue,
    citecolor=blue,
    urlcolor=magenta
]{hyperref}
\usepackage{array}

\begin{document}

\title{Channel Estimation for Movable Antenna Systems: Challenges, Solutions, and Opportunities}

\author{Linchu Chen, Zhendong Li, Zile Zou, Zhou Su, Lin Chen, Ruoyu Zhang, Qingqing Wu, and Wen Chen
\thanks{Linchu Chen, Zhendong Li, and Zile Zou are with the School of Information and Communication Engineering, Xi’an Jiaotong University, Xi’an 710049, China (email: chenlinchu@stu.xjtu.edu.cn; lizhendong@xjtu.edu.cn; zouzile@stu.xjtu.edu.cn). Zhou Su is with the School of Cyber Science and Engineering, Xi'an Jiaotong University, Xi'an 710049, China (email: zhousu@ieee.org). Lin Chen is with the Department of Electrical and Computer Engineering, Stevens Institute of Technology, Hoboken, NJ 07030, USA (e-mail: lchen53@stevens.edu). Ruoyu Zhang is with the School of Electronic and Optical Engineering, Nanjing University of Science and Technology, Nanjing 210094, China (e-mail: ryzhang19@njust.edu.cn). Qingqing Wu and Wen Chen are with the School of Integrated Circuits, Shanghai Jiao Tong University, Shanghai 200240, China (e-mail: qingqingwu@sjtu.edu.cn; wenchen@sjtu.edu.cn). (Corresponding author: Zhendong Li)}
\vspace{-1.5em}}

\maketitle

\begin{abstract}
Movable antenna (MA) has emerged as a promising technology for future wireless networks by exploiting channel variation over local antenna movement regions. However, accurate and efficient channel acquisition in MA systems remains challenging due to the trade-off between estimation accuracy and computational complexity. In this article, the MA channel model and the associated estimation framework are first reviewed, where channel information over the movement region is reconstructed from finite measurements by exploiting shared path parameters. The structured dependence of MA observations across space, time, and frequency naturally motivates the adoption of tensor-based modeling for channel estimation. Subsequently, we discuss the tensor-based signal model and corresponding parameter estimation methods from multidimensional observations. These methods are further compared with conventional channel estimation methods in terms of estimation accuracy, computational complexity, and general applicability. Furthermore, a representative case study is provided to illustrate the performance and characteristics of different algorithms under MA channel estimation settings. Finally, some future research directions for tensor decomposition-based channel estimation in MA systems are outlined.

\end{abstract}

\section{Introduction}
\IEEEPARstart{T}{he} multiple-input multiple-output (MIMO) technique substantially boosts the achievable throughput of wireless networks by exploiting an abundance of available spectrum and spatial diversity\cite{8284058}. Nevertheless, the inherent geometric constraints of conventional fixed antenna systems (FAS) limit the extent to which the available degrees of freedom (DoFs) in wireless channels can be fully utilized\cite{10236898}. To overcome these limitations, movable antenna (MA) is investigated as an efficient scheme to utilize the DoFs more efficiently\cite{10497534}. Compared with  {conventional FAS, which passively experience} random channel fading, MAs can intelligently adjust their positions to seek improved channel states. This capability allows MA systems to achieve enhanced performance while maintaining or reducing the number of antennas and radio frequency (RF) chains compared to conventional FAS.

However, such performance enhancement depends on rapid and accurate estimation of instantaneous channel parameters, including the angle-of-arrival (AoA), angle-of-departure (AoD), time delay, and frequency shift. In MA systems, the channel impulse response is inherently dependent on the antenna position coordinates\cite{10497534}, which fundamentally differentiates the channel estimation problem from that of conventional FAS. Specifically, from a structural perspective, the channel matrix in MA systems becomes a function of the antenna positions due to their mobility. Consequently, the spacing between adjacent antennas varies as the antenna positions change, and the resulting steering vector no longer exhibits a Vandermonde structure. Furthermore, from an algorithmic perspective, antenna position optimization in MA systems requires acquiring accurate channel state information (CSI), thereby imposing more stringent requirements on both estimation accuracy and computational efficiency. Classical estimation methods based on the compressed sensing (CS) require discrete grids and inevitably suffer from grid mismatch losses. Additionally, the computational burden of multidimensional search renders CS-based approaches impractical. 

Recent studies have investigated channel and parameter estimation for MA systems. In~\cite{10236898}, the field response information was recovered in three successive stages. The AoD and AoA were separately estimated with the orthogonal matching pursuit (OMP) algorithm from measurements taken as the MAs at the transmitter and receiver move. The final coefficient estimation therefore relied on the preceding AoD and AoA estimates.  {Moreover, a Cram\'er-Rao bound (CRB) guided optimization of the MA array positions was studied to improve sensing accuracy and the angular parameters of the targets were estimated by the multiple signal classification (MUSIC) in~\cite{10643473}.} Learning-based estimators have also been developed for MA systems in \cite{Jang2025Learning}.
 {However, the observations in MA systems collected across antenna positions, subcarriers and  time slots exhibit high-dimensional structure. When these observations are processed in vector or matrix form, this multilinear structure is not explicitly exploited, which prevents the estimation accuracy from being further improved. This motivates an algorithm that exploits the multidimensional structure of the received signals.}

 {To address the challenges arising from high dimensionality and real time processing in MA systems, tensor  algorithm was proposed to achieve improved performance\cite{tensor3,10403776,tensor1}.} This technique naturally integrates the spatial, time and frequency dimensions and preserves the multilinear structure of the channel. On the one hand, tensor decomposition exhibits identifiable uniqueness  under certain conditions\cite{6573422}, fundamentally avoiding the issue of parameter matching. On the other hand, by fully exploiting the structural information embedded in high-dimensional signals, tensor decomposition-based method can achieve higher estimation accuracy than conventional parameter estimation approaches. Therefore, investigating tensor decomposition-based parameter estimation in MA systems is of significant practical relevance.

Given that research on MA channel estimation algorithms is still in its nascent stage, and considering the aforementioned challenges, this article aims to provide a systematic overview of tensor decomposition-based channel estimation for MA systems. First, an MA channel model that accounts for antenna position variations is established, and the corresponding received observations are organized into a fourth-order tensor. Subsequently, we provide an in-depth analysis on the core algorithmic designs and quantitatively evaluate their advantages through a case study. Finally, open challenges and future research directions in this area are outlined.

\section{Channel Characteristics and Estimation Challenges}
MA systems introduce a position-dependent channel structure that differs from FAS, since the antenna positions at the  {base station (BS) and mobile station (MS)} can be adjusted within local movement regions. This section reviews the corresponding channel model and estimation framework, and then summarizes the main challenges caused by limited measurements, position-dependent channel variation, and multidimensional observation structures.
\begin{figure}[H]
    \centering
    \includegraphics[width=\columnwidth]{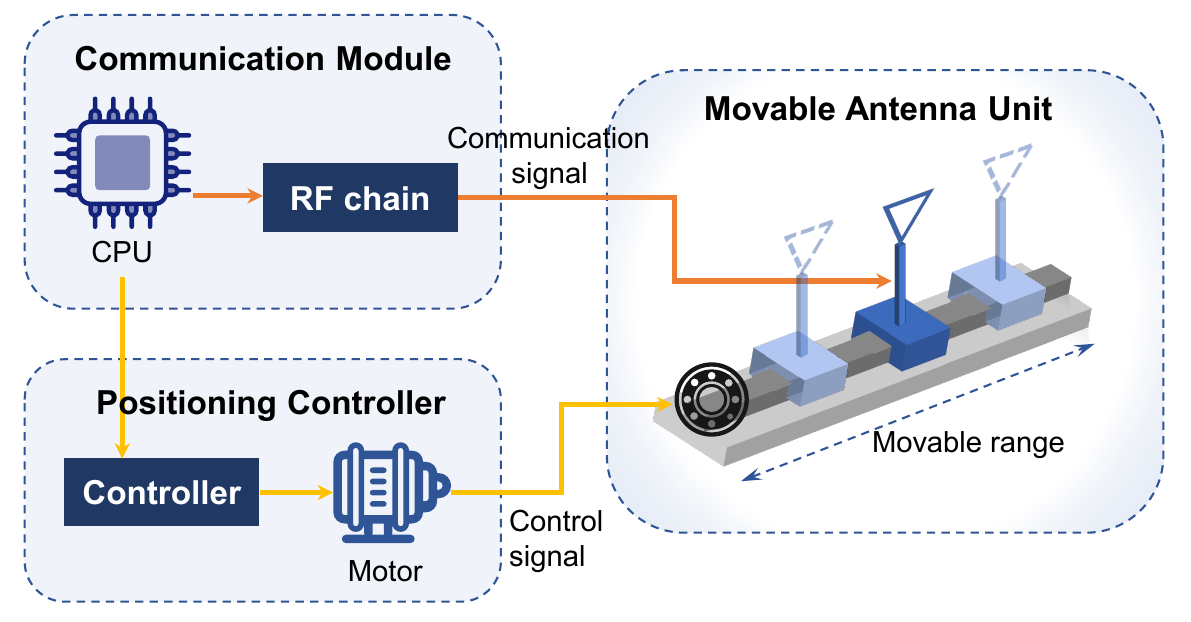}
    \caption{ {MA hardware architecture in which the RF-connected antenna is moved by a driving structure for position adjustment.}}
    \label{fig:ma_hardware_architecture}
\end{figure}
\subsection{Channel Characteristics}
\label{sec:channel}
MA system grants antenna adjustability to both the BS side and the MS side within their respective local movement regions. This local movement is enabled by the MA hardware architecture shown in Fig.~\ref{fig:ma_hardware_architecture}, where the MA units are integrated with RF chains and the associated driving structure for antenna position adjustment. From a practical implementation perspective, MA systems rely on external mechanical or electronic structures to convert control signaling and energy into mechanical motion, thereby enabling precise positioning of antenna elements within the prescribed local region~\cite{bajianle}.  {As the antenna position changes within the local movement region, the field responses at the BS and MS vary accordingly, which changes the phase contributions of different multipath components. Therefore, by adjusting the antenna positions, the deep fading problem caused by multipath effects can be avoided. This position flexibility creates additional spatial degrees of freedom that can improve subsequent communication performance by selecting favorable channel states.}

\begin{figure}
    \centering
    \includegraphics[width=\columnwidth]{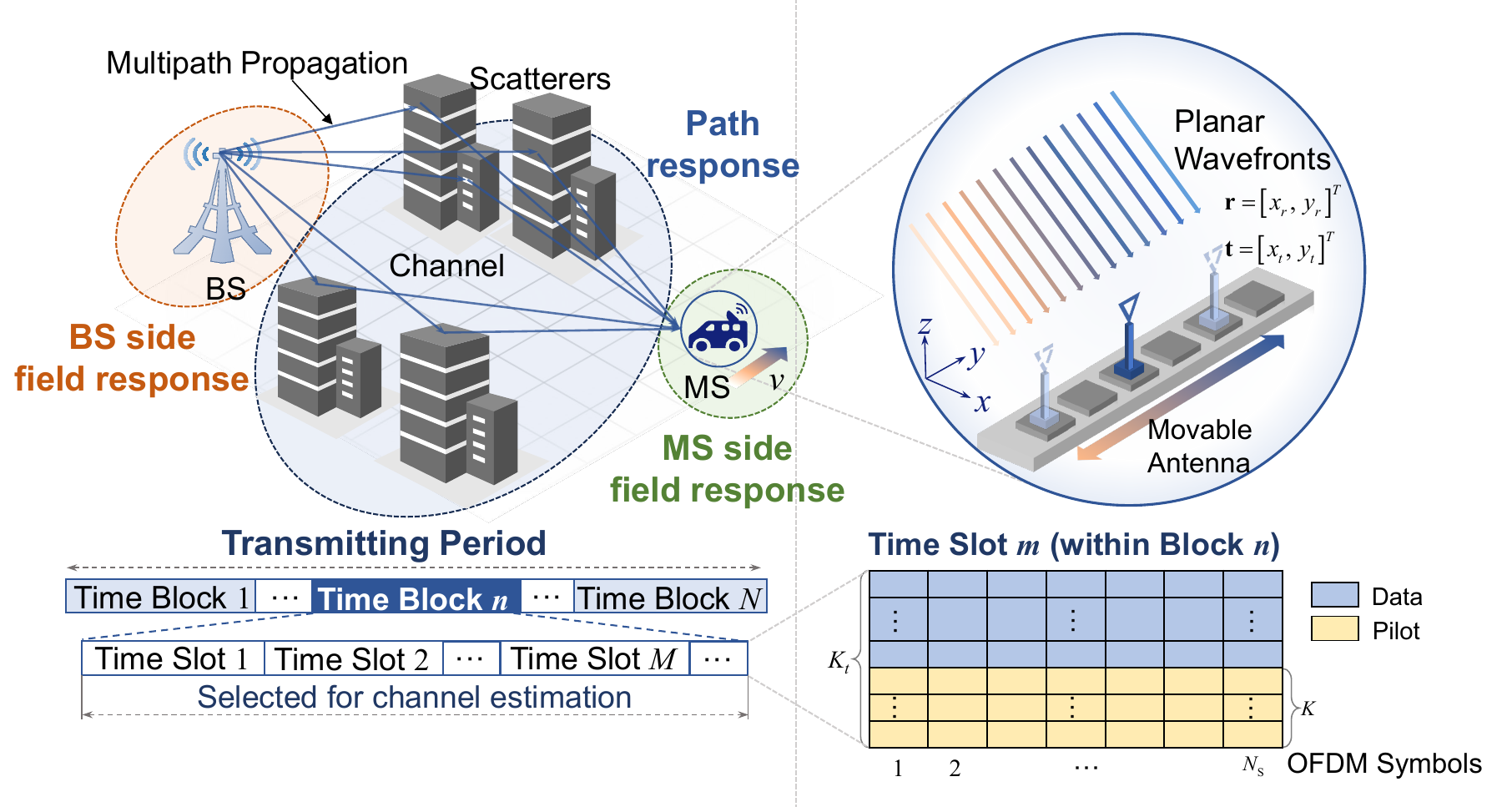} 
    \caption{  {MA system channel model and transmission frame structure design. The transmission period is divided into multiple time slots, and $K$ subcarriers are selected for channel estimation.}}
    \label{fig:field_path_coupling}
\end{figure}

The channel in MA systems differs from that in conventional FAS since it depends on the instantaneous antenna positions.  {Consider a BS with $N_{t}$ MA elements, an MS with $N_{r}$ MA elements and the channel can be written as $\mathbf{H}_{k,m}(\mathbf{r},\mathbf{t})\in \mathbb{C}^{N_{r}\times N_{t}}$, where $\mathbf{r}$ and $\mathbf{t}$ are the receive and transmit antenna position vectors, and $k=1,\ldots,K$ and $m=1,\ldots,M$ index the selected pilot subcarriers and the time slots within a time block. The channel is formed by $R$ propagation paths, the $r$-th of which has complex gain $\beta_{r}$,  AoA $\theta_{r}$,  AoD $\phi_{r}$, delay $\tau_{r}$ and Doppler shift $\nu_{r}$. The channel can be further written as $\mathbf{H}_{k,m}(\mathbf{r},\mathbf{t}) =\mathbf{F}(\mathbf{r},\boldsymbol{\Theta})\, \boldsymbol{\Sigma}_{k,m}\, \mathbf{G}(\mathbf{t},\boldsymbol{\Phi})^{T}$, where $\mathbf{F}(\mathbf{r},\boldsymbol{\Theta}) \in \mathbb{C}^{N_{r}\times R}$ and $\mathbf{G}(\mathbf{t},\boldsymbol{\Phi}) \in \mathbb{C}^{N_{t}\times R}$ denote the field response matrices, and their columns are beam steering vectors, e.g., $\mathbf{f}(\mathbf{r};\theta_r)=[e^{j\frac{2\pi}{\lambda}\rho(x^r_1; \theta_r)},\cdots,\ e^{j\frac{2\pi}{\lambda}\rho(x^r_{N_r}; \theta_r)}]^T$ in which $j\frac{2\pi}{\lambda}\rho(x; \theta) =j\frac{2\pi}{\lambda} x\text{cos}\theta$ denotes the phase determined by the antenna coordinates and the angle parameter, and the structure of $\mathbf{g}(\mathbf{t},{\phi}_r)$ is similar. The path response matrix $\boldsymbol{\Sigma}_{k,m} \in \mathbb{C}^{R\times R}$ is diagonal, with $r$-th entry $\beta_{r}e^{-j2\pi(k-1)\Delta f\tau_{r}}e^{j2\pi(m-1)T_{\mathrm{slot}}\nu_{r}}$, where $\Delta f$ is the subcarrier spacing and $T_{\mathrm{slot}}$ is the time  slot duration. The received signal is $\mathbf{Y}_{k,m} =\mathbf{Q}\big(\mathbf{H}_{k,m}(\mathbf{r},\mathbf{t})\mathbf{X} +\mathbf{N}_{k,m}\big)$, where $\mathbf{X} \in  \mathbb{C}^{N_{t}\times N_{s}}$ denotes the transmitting pilot matrix containing the distinct symbol information transmitted by each antenna, $\mathbf{Q} \in \mathbb{C}^{Q_{\text{MS}}\times N_{r}}$ denotes the combining matrix that maps the $N_{r}$ antenna outputs to the $Q_{\text{MS}}$ RF chains, $\mathbf{N}_{k,m}$ is the noise at the receive antennas. The field response matrices $\mathbf{F}(\mathbf{r},\boldsymbol{\Theta})$ and $\mathbf{G}(\mathbf{t},\boldsymbol{\Phi})$ are determined by the angular parameters and the position vectors $\mathbf{r}$ and $\mathbf{t}$, while $\boldsymbol{\Sigma}$ is governed by the gains ${\beta}_r$, delays ${\tau}_r$ and Doppler shifts ${\nu}_r$. These matrices carry the CSI of the MA system, and Fig.~\ref{fig:field_path_coupling} illustrates the channel and the transmission structure.} This representation characterizes the distinctive position dependent behavior of MA channels. Note that although antenna movement changes the position-dependent field responses, it does not create a new propagation environment.  {The channels at different position pairs are different observations of the same finite multipath environment and share the same AoA, AoD, delay and path gain~\cite{10403776}. Moving an antenna changes only the phase at which each path is received, not the path itself, so the observations at different positions are repeated samples of the same propagation parameters under different phases, with the same five parameters appearing in every observation.}

\subsection{Estimation Framework}
 {The objective of MA channel estimation is to reconstruct the corresponding channels among all candidate positions. In addition, more accurate parameter estimation can be achieved through antenna position optimization. The corresponding estimation framework can be summarized from the following key points. (i) Channel measurement at finite MA positions: Since directly measuring the channels over all candidate position pairs would introduce prohibitively large pilot overhead, practical MA channel estimation acquires channel measurements only at a limited number of selected MA positions. In practice, the channel information across MA candidate positions is not completely independent as mentioned in Section \ref{sec:channel}. Therefore, it is applicable to acquire channel measurements at a limited number of positions and then recover the full channel information from these samples. (ii) Regional channel reconstruction: The channels between candidate positions are determined by several sets of key parameters. If these key parameters can be estimated within a few measurements or even a single measurement, the complete channel between all transmit and receive ports can be directly reconstructed.  Moreover, the number of paths also needs to be estimated prior to parameter estimation. It can be estimated by algorithms based on the minimum description length (MDL) criterion \cite{7914672}. (iii) MA position optimization: To fully exploit the advantages of the MA system, more accurate parameter estimation can also be achieved through antenna position optimization. An optimization strategy can be adopted by optimizing the antenna positions at both the MS and the BS  by minimizing the CRB of estimation\cite{10643473}.  }

\subsection{Challenges}
\label{challenge}
In existing estimation frameworks, the difficulty arises from both observation modeling and parameter extraction.  {As mentioned above, if one directly estimates the channel for every port pair, the resulting estimation overhead would be prohibitively large. It is noted that the transmitted signal exhibits sparsity in domains such as angle and delay, and its sparsity level equals the number of paths $R$, since the signal propagates only along a few dominant paths. Therefore, the received signal can be represented as a superposition of signals over these sparse paths. By estimating the key parameters of these paths, the channel state information corresponding to each port pair can be recovered. This imposes a requirement on the parameter estimation algorithm that it should properly exploit the sparsity of the received signal.}

 {Moreover, the channel varies across different antennas, subcarriers, and time slots. By stacking $\mathbf{Y}_{k,m}$ over these dimensions, one obtains a high-dimensional signal $\mathcal{Y}$ that exhibits distinct characteristics in the time, frequency, and spatial domains.  {Conventional parameter estimation algorithms, such as MUSIC and OMP, flatten the received signal into a matrix form, which leaves the multilinear structure among the four dimensions implicit and therefore not explicitly exploited. Such insufficient exploitation of the multilinear coupling leads to a degradation in estimation accuracy.}}

\section{{Tensor-Based Channel Estimation Framework}}
\label{sec:tensor}
 {Section~\ref{sec:channel} showed that the observations collected at different antenna positions, subcarriers and slots are generated by the same $R$ propagation paths. The received signal is therefore a sum of $R$ rank-one terms and admits a rank-$R$ canonical polyadic (CP) representation. Since only $R$ dominant paths contribute, the tensor has CP rank $R$ under distinct path parameters and can be expressed as $\mathcal{Y}\approx\sum_{r=1}^{R}\beta_{r}\mathbf{a}_{r}\circ\mathbf{b}_{r}\circ\mathbf{c}_{r}\circ\mathbf{d}_{r}$, where $\circ$ denotes the vector outer product, and $\mathbf{a}_{r},\mathbf{b}_{r},\mathbf{c}_{r},\mathbf{d}_{r}$ are the $r$-th columns of the four factor matrices. Specifically, $\mathbf{a}_{r}\in\mathbb{C}^{Q_{\text{MS}}}$ is the combined receive array response for the AoA $\theta_{r}$, $\mathbf{b}_{r}\in\mathbb{C}^{K}$ contains the phases set by the delay $\tau_{r}$, $\mathbf{c}_{r}\in\mathbb{C}^{M}$ contains the phases set by the Doppler shift $\nu_{r}$, and $\mathbf{d}_{r}\in\mathbb{C}^{N_{s}}$ is the pilot-coded transmit array response for the AoD $\phi_{r}$. The four factor matrices are $\mathbf{A}=\mathbf{Q}\mathbf{F}(\mathbf{r},\boldsymbol{\Theta})\in\mathbb{C}^{Q_{\text{MS}}\times R}$, $\mathbf{B}\in\mathbb{C}^{K\times R}$ with $[\mathbf{B}]_{k,r}=e^{-j2\pi(k-1)\Delta f\tau_{r}}$, $\mathbf{C}\in\mathbb{C}^{M\times R}$ with $[\mathbf{C}]_{m,r}=e^{j2\pi(m-1)T_{\mathrm{slot}}\nu_{r}}$, and $\mathbf{D}=\mathbf{X}^{T}\mathbf{G}(\mathbf{t},\boldsymbol{\Phi})\in\mathbb{C}^{N_{s}\times R}$. They correspond to the receive spatial, frequency, slot and transmit symbol dimensions, respectively. Decomposing the tensor recovers these factors, and exploiting their structure in parameter estimation improves the accuracy.}  Fig.~\ref{fig:flow} summarizes the tensor decomposition-based framework, where the observations are decomposed through alternating least squares (ALS) or structured canonical polyadic decomposition (SCPD), and the recovered factors are then used for physical parameter extraction and channel reconstruction over the movement region.

\subsection{Alternating Least Squares}
Based on the CP decomposition model, ALS estimates the factor matrices in an iterative manner. As illustrated in Fig. ~\ref{fig:flow}, the ALS-based recovery updates one factor matrix at a time while fixing the remaining factors, thereby transforming the original tensor decomposition problem into a sequence of LS subproblems. ALS is a classical algorithm for computing the CP decomposition of received tensor signals.  {Its core idea is to decompose a high-dimensional nonlinear low-rank tensor approximation problem into a series of linear LS subproblems that can be solved in closed form by alternately fixing the other factors~\cite{tensor1}.} The essence of this alternating iteration is a cyclic LS projection among the  {four} factor spaces, where each update monotonically decreases the overall fitting error. From an algebraic subspace perspective, ALS employs a divide-and-solve strategy in which the algorithm focuses on the column space spanned by one factor matrix. In this way, the original problem is projected onto the matricized observation along the corresponding mode and formulated as a standard linear regression problem. From a convergence viewpoint, ALS is a block coordinate descent optimization method that solves one subproblem at each step.  Overall, ALS provides a practical framework that breaks a complex tensor decomposition problem into repeatedly executed linear-algebra subproblems, and in scenarios without special structural constraints, it is one of the most direct and flexible tools for tensor decomposition.

%The basic idea is to alternatively minimizes the data fitting error with respect to one of the factor matrices, while keeping the other factor matrices fixed. Then solve the resulting LS subproblem in each iteration and the iterations continue until the fitting error converges or a prescribed stopping criterion is reached. A practical feature of ALS is that it imposes no explicit structural constraint on the factor matrices during the iterative updates~\cite{7891546}. This makes it broadly applicable to CP decomposition problems. However, the specific structure contained in the signal model is not explicitly exploited in the estimation stage, and the decomposition may also be sensitive to initialization and observation quality, especially in the presence of noise or other estimation errors. As a result, ALS may suffer from performance degradation in practical MA channel estimation scenarios.
\iffalse
\begin{figure*}[ht] 
    \centering 
    \includegraphics[width=\textwidth]{fig/flow.pdf} 
    \caption{Tensor decomposition-based channel estimation framework.} \label{fig:flow} 
\end{figure*} 
\fi

\begin{figure}
    \centering
    \includegraphics[width=1.05\columnwidth]{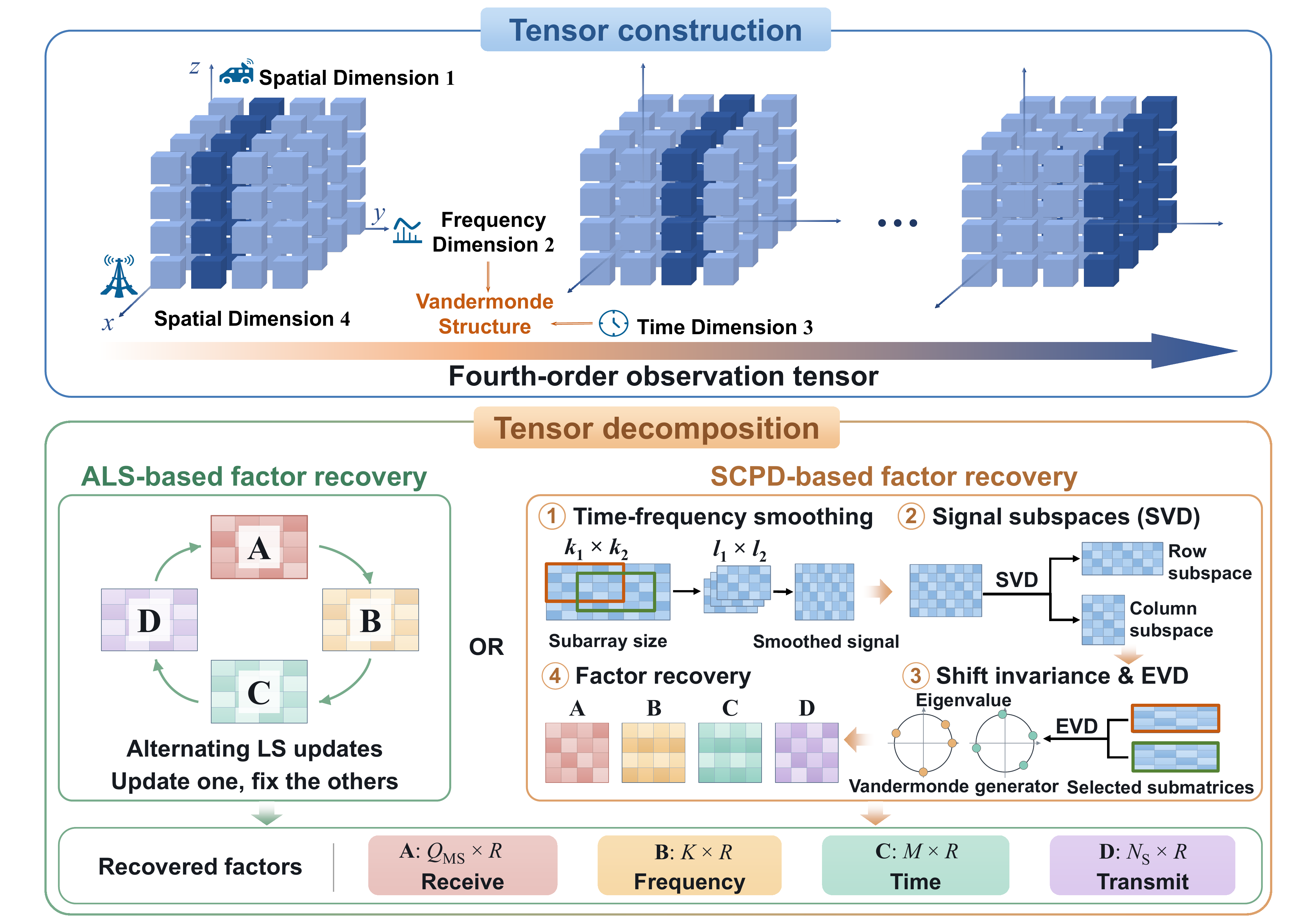}
    \caption{ {Tensor construction and factor decomposition. Received signals are stacked over slots, subcarriers, RF chains and symbols into a fourth-order tensor, which is then decomposed by ALS or SCPD.}}
    \label{fig:flow}
\end{figure}

\subsection{Structured Canonical Polyadic Decomposition}
\label{sec:scpd}
 {While ALS provides a general iterative mechanism for factor matrix estimation, it does not explicitly exploit the structure contained in the MA signal model. In the received signal, the frequency spacing between adjacent subcarriers and the time interval between successive slots are uniform. As shown in Section \ref{sec:channel}, this uniformity makes the phase of the entries in each column of the frequency and slot factors change by a fixed amount from one row to the next, with the increment set by the corresponding path parameter. The frequency factor is therefore determined by the path delays, and the slot factor by the Doppler shifts. Both are Vandermonde, which provides the basis for an alternative tensor decomposition method and motivates the use of SCPD.} As summarized in Fig.~\ref{fig:flow}, this structured recovery process is carried out by explicitly exploiting these structured factor matrices rather than treating all factors as generic unknown matrices\cite{6573422}. The exploitation of Vandermonde structure in the factor matrices manifests itself in two aspects. First, leveraging this structure enables the use of spatial smoothing techniques to resolve the potential rank deficiency problem in the factor matrices, which would otherwise severely compromise the accuracy of the estimation. Second,  {this structure also allows for the direct derivation of a closed-form solution for each factor matrix}.

Specifically, SCPD reformulates structured tensor decomposition as a similarity transformation problem in the signal subspace of a suitably constructed data matrix.  {To this end, spatial smoothing is applied along the frequency and time modes, whose factor matrices exhibit Vandermonde structures. Let \(k_1\) and \(k_2\) denote the smoothing-window lengths along these two modes, respectively, and let \(l_1=K-k_1+1\) and \(l_2=M-k_2+1\) denote the corresponding numbers of overlapping shifts. The resulting \(l_1l_2\) shifted data blocks are concatenated into a smoothed matrix of size \(k_1k_2Q_{\mathrm{MS}}\times l_1l_2N_s\). By exploiting the shared path structure across these blocks, smoothing can alleviate rank deficiency and enable structured factor recovery when the required rank conditions are satisfied.} Singular value decomposition (SVD) is then applied to extract an orthonormal basis for the signal subspace from the \(R\) dominant left singular vectors. This basis spans the same column space as the corresponding structured factors and is related to them through an unknown nonsingular mixing matrix. Exploiting the Vandermonde shift-invariance property, two appropriately row-selected submatrices of this basis are used to construct a matrix that is similar to a diagonal matrix of Vandermonde generators. Eigenvalue decomposition (EVD) therefore recovers these generators as eigenvalues, enabling algebraic reconstruction of the structured factors without iterative nonlinear optimization. The remaining non-Vandermonde factors are subsequently recovered through the Khatri-Rao product structure, where the Khatri–Rao product denotes the column-wise Kronecker product. This procedure converts the structural information in the MA channel model into a sequence of smoothing, subspace extraction, and algebraic factor-recovery operations~\cite{6573422}.

\subsection{Parameter Extraction}
 {Parameter extraction maps the recovered factor matrices to the physical attributes of each propagation path. The frequency and time factors reveal path delay and frequency offset through their phase evolution, while the spatial factors provide AoA and AoD through their match with the corresponding field responses. Once these parameters are determined, the path gains can be estimated for channel reconstruction.} From an algebraic subspace perspective, the extraction of physical parameters from the factor matrices depends on the structural characteristics of each dimension. For the frequency and time dimensions, the factor matrices naturally exhibit a Vandermonde structure, which embodies a shift-invariance property. This endows the column space spanned by these matrices with self-similarity, enabling the estimation of time delays and frequency offsets to be transformed into an eigenvalue problem via a similarity transformation between subspaces, thereby yielding a closed-form solution.  {As discussed in Section~\ref{sec:scpd}, the equally spaced subcarriers and slots give the frequency and time factors a Vandermonde structure, so the delays and frequency shifts are recovered by estimating the per-column phase increment.} In contrast, the spatial factor matrices lose any Vandermonde or similar structural constraints due to the MA antenna array architecture and their coupling with the combiner and the precoding matrix. Consequently, angular parameters cannot be directly extracted through subspace transformations. Instead, we can resort to a parametric matching framework in which a correlation-based search projects the observed spatial subspace coordinates onto a set of predefined array response vectors to perform peak-finding, thereby estimating the angles of arrival and departure. Once the physical parameters are determined, the estimation of path gains reduces to a linear least-squares problem.

\section{Advantages of Tensor Decomposition-Based Channel Estimation Algorithm}
This section discusses the advantages of tensor-based channel estimation algorithm for MA systems from three perspectives, namely estimation accuracy, computational complexity, and applicability. The following discussion compares tensor-based methods with conventional estimation algorithms and also discusses the difference between ALS and SCPD.

% Required packages:
% \usepackage{array,booktabs,multirow,amsmath}

% Required packages:
% \usepackage{array,booktabs,multirow,amsmath}

\begin{table*}[!t]
\caption{Comparison of representative channel parameter estimation
algorithms for MA systems.}
\label{tab:comparison_methods}
\centering
\footnotesize
\renewcommand{\arraystretch}{1.2}
\setlength{\tabcolsep}{3pt}

\begin{tabular}{@{}
    >{\centering\arraybackslash}m{1.5cm}
    >{\centering\arraybackslash}m{1.5cm}
    >{\raggedright\arraybackslash}m{3.7cm}
    >{\raggedright\arraybackslash}m{3.8cm}
    >{\raggedright\arraybackslash}m{3.8cm}@{}}
\toprule
\textbf{Category}
& \textbf{Algorithm}
& \textbf{Accuracy  characteristics}
& \textbf{Dominant total complexity}
& \textbf{Applicability} \\
\midrule

\multirow[c]{2}{*}[-1.5\baselineskip]{%
    \shortstack{\textbf{Tensor-}\\\textbf{based}}}
& ALS
& High accuracy; sensitive to initialization and convergence.
& $\mathcal{O}\!\left(
N_{\rm iter}Q_{\rm MS}KMN_sR
\right)$
& General CP models satisfying identifiability conditions. \\
\cmidrule(lr){2-5}

& SCPD
& High accuracy through exploitation of Vandermonde structure.
& $\mathcal{O}\!\left(
k_1k_2l_1l_2KMQ_{\rm MS}^{2}N_s
\right)$
& Requires Vandermonde structure and sufficient
smoothed-data rank. \\
\midrule

\shortstack{\textbf{Subspace-}\\\textbf{based}}
& MUSIC
& Sensitive to SNR, source separation, and subspace estimation.
& $\mathcal{O}\!\left(
G^2(Q_{\rm MS}+K+M+N_s)
\right)$
& Requires separable signal and noise subspaces. \\
\midrule

\shortstack{\textbf{CS-}\\\textbf{based}}
& OMP
& Limited by grid mismatch and dictionary coherence.
& $\mathcal{O}\!\left(
RGQ_{\rm MS}KMN_s
\right)$
& Requires a suitable sparse dictionary. \\

\bottomrule
\end{tabular}
\end{table*}

\subsection{Estimation Accuracy}
An important advantage of tensor-based channel estimation is that it organizes the observations into a higher-order tensor, retaining the coupling across dimensions that conventional one-dimensional processing discards. Consistent with this structural advantage, tensor decomposition-based algorithm can achieve higher accuracy than MUSIC and OMP in channel parameter estimation. In particular,  {the OMP algorithm requires constructing an overcomplete dictionary and performing iterative atom selection\cite{10497534}.  At each iteration, it selects from the dictionary the atom that is most correlated with the column vector of the received signal, and its estimation accuracy is limited by the number of search grid points. Moreover, the errors incurred during the iterative steps will propagate and accumulate, ultimately resulting in a significant estimation error.} MUSIC is sensitive to the separation of spectral peaks, and its performance degrades when closely spaced peaks become difficult to resolve. These limitations indicate that MUSIC and OMP are more sensitive to noise and resolution conditions, whereas tensor decomposition-based algorithms such as ALS and SCPD can achieve more accurate parameter estimation by exploiting this multidimensional structure.  {Comparing the two tensor decomposition algorithms under consideration, SCPD utilizes the structural information inherent in the high-dimensional signal, whereas ALS neglects the intrinsic structure across all factor matrices. Consequently, SCPD demonstrates superior estimation accuracy and algorithmic stability.}

\subsection{Computational Complexity}
 {The complexity analyses of several algorithms are as follows.  For ALS, the processing pipeline consists of SVD initialization, alternating LS updates, and physical parameter extraction. The dominant overhead stems from the tensor and Khatri-Rao product multiplications over \(N_{\rm iter}\) full update rounds. Upon accounting for the angular grid matching, the dominant expression of the total complexity is given by \(\mathcal{O}(N_{\rm iter}Q_{\rm MS}N_sKMR+RG(Q_{\rm MS}+N_s))\). Extracting the delay and Doppler parameters via phase relationships further requires \(\mathcal{O}(R(K+M))\) operations. When the repeated LS updates dominate, the total complexity can be simplified to \(\mathcal{O}(N_{\rm iter}Q_{\rm MS}N_sKMR)\). For SCPD, the processing pipeline involves smoothed matrix construction, truncated SVD, structured factor recovery, and physical parameter extraction. The dominant overhead originates from the matrix multiplications required by the dense smoothing operations corresponding to the \(l_1l_2\) overlapping shifts. The dominant expression of the total complexity is \(\mathcal{O}(k_1k_2l_1l_2KMQ_{\rm MS}^2N_s+RG(Q_{\rm MS}+N_s))\), while the extraction of delay and Doppler parameters based on phase relationships further requires \(\mathcal{O}(R(K+M))\) operations. When the dense smoothing operations dominate, this can be simplified to \(\mathcal{O}(k_1k_2l_1l_2KMQ_{\rm MS}^2N_s)\). For MUSIC, the processing pipeline includes covariance estimation,  EVD, spectral search, and path gain estimation. The dominant expression of the total complexity is given by \(\mathcal{O}(Q_{\rm MS}N_sKM(Q_{\rm MS}+N_s+K+M)+G^2(Q_{\rm MS}+K+M+N_s))\), in which the spectral search dominates, leading to a simplified expression of \(\mathcal{O}(G^2(Q_{\rm MS}+K+M+N_s))\). For OMP, the processing pipeline involves dictionary construction, \(R\) iterations of atom selection across each mode, residual updating, and path gain estimation. The dominant overhead arises from the repetitive computation of the correlation between the dictionary and the residual. Under low-rank conditions, the dominant expression of the total complexity is \(\mathcal{O}(RQ_{\rm MS}N_sKM(G+Q_{\rm MS}+N_s+K+M))\). When the grid size is sufficiently large and the correlation computation dominates, it can be simplified to \(\mathcal{O}(RGQ_{\rm MS}N_sKM)\). Here, \(G\) denotes the number of grid points in a representative one-dimensional search.}

 {The computational advantage of tensor-based methods is closely related to their structured processing for MA channel estimation. Compared with OMP, tensor-based methods avoid the repeated multiplication between the sparse dictionary matrix and the received signal matrix during iterative atom selection. They also avoid the subspace decomposition and subsequent spectral search required by MUSIC for parameter estimation. Instead, the received signals are organized into a structured tensor, from which the factor matrices are recovered first and then used for parameter extraction.  {ALS and SCPD have different computational behavior because they recover the factor matrices in different ways. ALS performs successive LS updates directly on tensor unfoldings and does not explicitly construct spatially smoothed data, so its cost is governed mainly by the original tensor dimensions. In contrast, SCPD forms a spatially smoothed matrix by repeatedly applying selection matrices to the unfolded observation matrix for multiple overlapping subarrays. As the factor matrix dimensions increase, both the number of such selection matrix multiplications and the sizes of the resulting submatrices increase. Spatial smoothing therefore becomes the main computational bottleneck at large dimensions. When the factor matrix dimensions are small or moderate, however, these matrices remain compact and the algebraic SCPD recovery can avoid the repeated LS updates required by ALS.} The choice between ALS and SCPD therefore depends on the scale of the received tensor and the desired balance between efficiency and structural exploitation.}

\subsection{General Applicability}
From the perspective of applicability, tensor decomposition-based methods differ from conventional parameter estimation methods in how they handle the received observations. By rearranging multiple received signal vectors or matrices into a multidimensional tensor, they can work directly with the structured observation model of MA systems. In this sense, they are less tied to discretized sparse representations than OMP-based methods and less dependent on subspace separation and peak-resolution conditions than MUSIC-based methods.  {Furthermore, when the scenario changes, it is sufficient to adjust the structures of the corresponding factor matrices, and the tensor decomposition algorithm remains directly applicable. For example, the structure of the factor matrix along the frequency dimension can be adapted to accommodate scenarios with limited spectrum resources, or the structure of the factor matrix along the spatial dimension can be modified to extend the proposed algorithm to a wider variety of movable antenna systems, such as rotatable antennas, fluid antennas, and pinching antennas~\cite{10883029}.}

Within the tensor-decomposition framework, successful channel reconstruction requires uniqueness of the underlying decomposition.  {The uniqueness condition is of critical importance for tensor decomposition algorithms, as it pertains to whether the structural information contained in the factor matrices uniquely corresponds to the received signal. Specifically, the uniqueness condition requires that the Kruskal rank of each factor matrix,  {defined as the largest integer $k$ such that any $k$ columns of the matrix are linearly independent}, satisfy $k_{\mathbf{A}} + k_{\mathbf{B}} + k_{\mathbf{C}} + k_{\mathbf{D}} \geq 2R + 3$.  {Satisfying the uniqueness condition guarantees that the decomposition results are unique up to the inherent permutation and scaling ambiguities, thereby matching the true physical model and ensuring the validity of the tensor decomposition-based parameter estimation method} ~\cite{6573422}.}  {In MA systems, the spatial-domain information can be adjusted by altering the antenna array geometry, which implies that larger $k_\mathbf{A}$ and $k_\mathbf{D}$ can be achieved, thereby facilitating the satisfaction of the uniqueness condition.} 

ALS exhibits stronger applicability because it does not impose specific structural requirements on the factor matrices and can be applied to general CP decomposition problems. In addition, each LS update can be adapted into a constrained LS formulation, which makes ALS suitable for more flexible estimation settings. By contrast, SCPD is more closely tied to structured factors and relies on spatial smoothing together with subspace-based recovery. Accordingly, ALS is more flexible from the viewpoint of applicability, whereas SCPD is more structure-dependent but remains effective when the required model structure is available. The main differences among representative channel parameter estimation methods for MA systems are summarized in Table~\ref{tab:comparison_methods}.

\section{Case Study}
%Building on the preceding discussion regarding the tensor decomposition based channel parameter estimation framework in MA systems, this section utilizes resource allocation in an MA wireless network as a case study. It demonstrates the solution approach and presents the results of applying GML to address practical resource allocation problems in MA networks.
To demonstrate the superior performance of tensor decomposition-based parameter estimation algorithms in MA systems, this section presents a comprehensive evaluation of various channel estimation algorithms, encompassing both estimation accuracy and computational complexity. The simulation considers a downlink MA system, where the BS and MS are each equipped with a linear MA antenna array consisting of \text{10} elements.  {During the several time slots used for estimation, the antenna coordinates are randomly generated within the range of candidate positions.} The number of RF chains is set to be $\text{7}$. The number of candidate positions is \text{32}. A time block contains \text{8} time slots, and each time slot comprises \text{10} orthogonal frequency division multiplexing (OFDM) symbols. The transmission bandwidth is set to \text{10} MHz and is divided into \text{128} subcarriers, \text{16} of which are used for parameter estimation. The carrier frequency is set to \text{30} GHz, and the precoding and combining matrices are randomly generated.  {The number of paths is set to be $\text{3}$, and all channel parameters are uniformly and randomly generated within reasonable ranges.} Specifically, the delay parameters range from $\text{0.1}$ $\mu$s to $\text{0.6}$ $\mu$s, the angle parameters are within $\text{60}^\circ$ to $\text{120}^\circ$, and the maximum Doppler shift is $\text{2}$ kHz.  {The ALS algorithm is initialized using a combination of random initialization and truncated SVD initialization and the stopping criterion for the ALS algorithm is either the normalized squared reconstruction residual falling below $\text{10}^\text{{-5}}$ or the maximum number of iterations $N_\text{iter}=\text{60}$ being reached.} MUSIC and OMP are selected as benchmarks and the number of grid points for the two comparison algorithms is set to 1200. The number of Monte Carlo trials is set to 500.  {The simulation results were conducted using MATLAB R2024b Update 2 on a $\text{64}$-bit Windows workstation equipped with an Intel Core Ultra 7 265K processor and approximately $\text{32}$ GB of RAM.}
\begin{figure}
    \centering
    \includegraphics[width=0.9\linewidth]{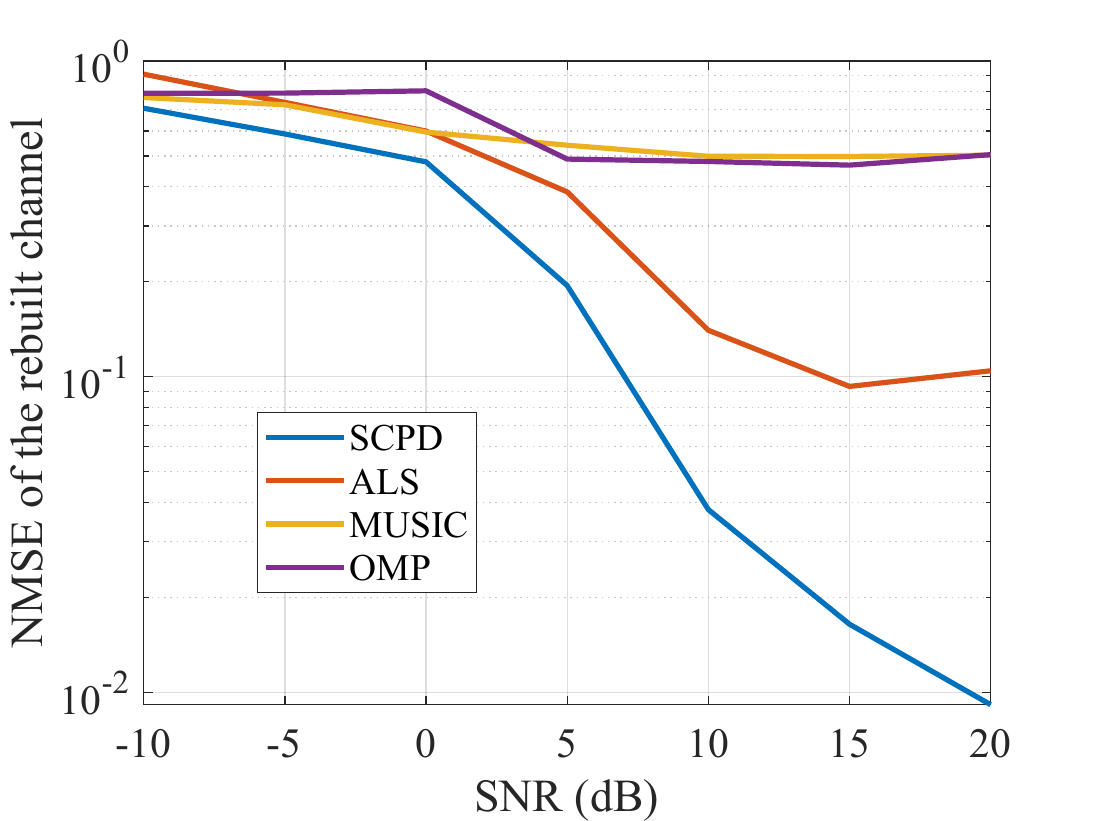}
    \caption{ {NMSE of the rebuilt MA channel versus SNR for tensor-based and conventional estimation algorithms.}}
    \label{nmse}
\end{figure}

Fig.  \ref{nmse} presents the channel estimation errors of several representative algorithms. Each algorithm first estimates channel parameters, including AoA, AoD, Doppler shift, delay, and channel gain. These parameters are then used to reconstruct the channel matrix, and the accuracy of each algorithm is evaluated by comparing the reconstructed channel with the ground-truth channel. It can be observed that as the signal-to-noise ratio (SNR) increases, the normalized mean squared error (NMSE) of the SCPD algorithm decreases, demonstrating the best estimation accuracy. In contrast, OMP and MUSIC exhibit inferior performance. This is because the successful channel reconstruction by OMP algorithm requires the system model to satisfy the restricted isometry property (RIP) condition. If this condition is not met, OMP will fail. And since factor matrices of the antenna dimension are not full column rank, the orthogonality between the noise subspace and the signal subspace is compromised, thereby causing the MUSIC algorithm to fail. This reflects that both algorithms impose strict requirements on the scenario. In contrast, the tensor decomposition-based algorithm is not restricted by the RIP condition and  {is robust against noise}. This approach fully exploits the structural information of the high-dimensional received signals and decomposes different factor matrices used for separated parameter estimation. By avoiding joint parameter estimation, it significantly enhances estimation accuracy.

\begin{figure}
    \centering
    \includegraphics[width=0.9\linewidth]{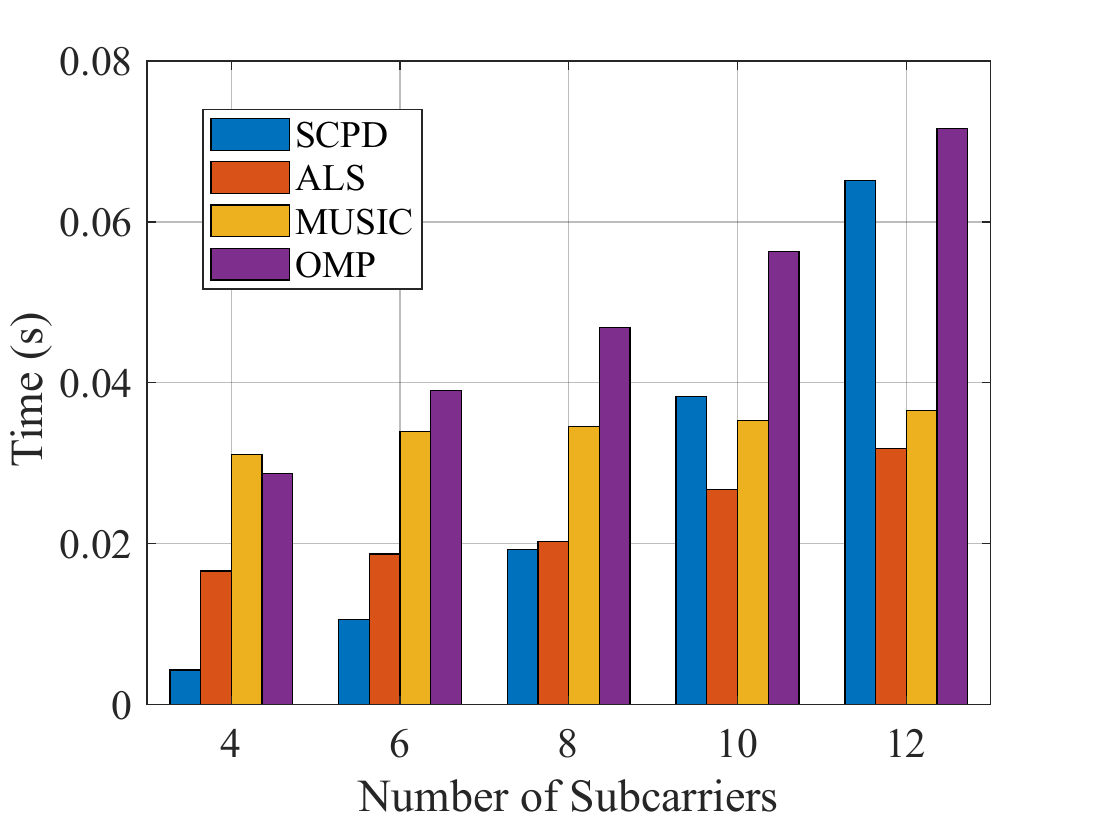}
    \caption{ {Runtime versus the number of subcarriers for tensor-based and conventional  estimation algorithms.}}
    \label{time}
\end{figure}

Fig.  \ref{time} illustrates the variation in algorithm runtime as the number of subcarriers used for channel estimation increases. An increase in the number of subcarriers leads to a larger size of the frequency-domain factor matrix in tensor decomposition. It can be observed that  {the ALS algorithm exhibits the lowest runtime in most scenarios}. The runtimes of MUSIC and OMP algorithms increase slightly with the number of subcarriers, whereas the runtime of the SCPD algorithm increases rapidly. For MUSIC and OMP algorithms, sufficiently large number of searching grids $G$ needs to be constructed, which primarily affect the runtime. Therefore, even when the dimension of the received signal is relatively small, i.e., $K$ = $\text{4}$, both algorithms still exhibit high computational time. %For SCPD algorithm, spatial smoothing operation constitutes the primary computational bottleneck as its complexity scales cubically with the dimension of the factor matrix. Therefore, when the dimension of the increases, the runtime grows rapidly due to the heavy overhead of spatial smoothing. Conversely, the SCPD algorithm achieves the shortest runtime, when the dimension of the factor matrix is small.
For the SCPD algorithm, spatial smoothing constitutes the primary computational bottleneck because its complexity scales cubically with the factor matrix dimension.  {When the factor matrix dimensions are small, the overlapping subarrays formed by spatial smoothing and the matrices used in the subsequent subspace operations remain compact. SCPD can then recover the structured factors through one algebraic procedure based on SVD and EVD, avoiding the repeated LS updates required by ALS. It can therefore achieve the shortest runtime in this small-dimensional regime. As the factor matrix dimensions increase, the enlarged spatial smoothing operation becomes dominant and the SCPD runtime grows rapidly.}

\section{Open Issues and Future Directions}
In this section, we propose several key future directions on tensor decomposition-based algorithm for MA systems. Existing research on tensor decomposition-based channel parameter estimation for MA systems has established a solid foundation. However, numerous opportunities remain to be explored for practical deployment and integration with emerging 6G paradigms. Future research can be pursued along two dimensions: the deepening of methodological approaches and the expansion of application scenarios.

\subsection{Deepening of Methodological Approaches}
Although the current CP decomposition algorithm based on ALS is widely used, there remains room for improvement in both its convergence speed and estimation accuracy. Meanwhile, the SCPD algorithm suffers from a rapid increase in computational complexity as dimensionality grows, due to the required spatial smoothing operation. 
A promising direction lies in integrating SCPD with ALS. Specifically, certain factor matrices of the received signal tensor naturally exhibit Vandermonde structures. By explicitly incorporating these structural constraints into each iteration of ALS, the feasible solution space is reduced, guiding the iterations toward the true parameter values more rapidly and thereby significantly decreasing the number of iterations needed for convergence. Moreover, the phase linear relationship inherent in the Vandermonde structure effectively mitigates noise perturbations, enabling the decomposed factor matrices to reflect the physical channel parameters more accurately. Consequently, this approach also enhances decomposition precision. Furthermore, future research can explore combinations of different structural constraints to accommodate a wider range of channel propagation characteristics.

%Current research primarily operates under ideal assumptions. Future work should focus on overcoming the non-idealities inherent in practical hardware scenarios. This includes establishing accurate tensor signal models that account for constraints such as positioning quantization errors, mechanical response delays, channel mismatch, and limited movement ranges, as well as designing corresponding robust decomposition algorithms. In addition, estimation algorithms can be integrated with dynamic channel tracking. To address channel time-variations caused by user mobility or environmental changes, adaptive tensor decomposition algorithms can be developed to enable low-overhead continuous parameter tracking and position servo control, rather than relying on independent snapshot-based estimation, thereby better aligning with practical scenario requirements.

\subsection{Expansion of Application Scenarios}
 {The tensor decomposition framework presented in this article can also be applied to integrated sensing and communication (ISAC). MA is attractive for ISAC because the antenna positions provide an additional degree of freedom. Moving the antennas changes the array response and can improve sensing accuracy and angular resolution without adding RF chains, and antenna position optimization has been shown to improve both sensing and communication performance in MA ISAC systems~\cite{ma2026movable}. Tensor methods are valuable for a complementary reason. The communication channel and the sensing echoes are both superpositions of a few components, and each component factorizes into angle, delay and Doppler terms. The received ISAC signal therefore has a multilinear structure that can be represented by a low-rank tensor, and the same decomposition recovers the channel and target parameters. This structure brings three benefits. First, it enables a unified formulation of channel estimation and target sensing with shared time-frequency resources, which reduces training overhead while improving estimation accuracy and sensing resolution~\cite{10403776}. Second, it reveals the common multilinear structure and the partially shared physical parameters between sensing and communication, so sensing information can be used to improve channel estimation~\cite{liang2026joint}. Third, full-order tensor models exploit all signal dimensions and support channel reconstruction with reduced pilot overhead, together with generic uniqueness conditions~\cite{he2025full}. Extending the framework of this article to MA ISAC is therefore promising. A unified received tensor can incorporate both communication signals and sensing echoes, and structured decomposition can extract communication multipath parameters and target distance, angle and velocity jointly.} Moreover, as antennas evolve toward higher frequency bands and larger apertures, near-field propagation effects become significant, rendering the traditional far-field plane-wave assumption invalid. Future research scenarios also include extremely large-scale and near-field MA systems. This necessitates the construction of near-field tensor models that incorporate spherical wavefronts and distance-dependent path loss. Finally, several assumptions in this article, such as movement time and positioning accuracy, remain idealized and deserve further study.

\section{Conclusion}
This article investigated tensor decomposition-based channel estimation for MA systems. By exploiting the shared path structure among different MA position pairs, the channel over the movement region can be reconstructed from finite measurements rather than measurements over all candidate position pairs. The received observations across antenna position, frequency, and time dimensions can be organized into a tensor form, which supports factor matrix recovery through ALS or SCPD, followed by physical parameter extraction and channel reconstruction. Compared with representative conventional algorithms, tensor decomposition-based algorithm exhibits favorable accuracy and complexity characteristics in the considered setting. Finally, open issues and future directions are discussed from the perspectives of methodological development and application expansion.

\balance
\bibliographystyle{IEEEtran}
\bibliography{reference}

\end{document}